\documentclass[reprint,superscriptaddress]{revtex4-1}
\usepackage{amssymb}
\usepackage{amsmath}
\usepackage{hyperref}
\usepackage{graphicx}
\usepackage{url} 
\usepackage{outline}
\usepackage{enumitem}
\usepackage{mhchem}
\usepackage{gensymb}
\usepackage[export]{adjustbox}
\usepackage{array}
\usepackage{braket}

\usepackage{listings}

\usepackage{pdfpages}

\makeatletter
\AtBeginDocument{\let\LS@rot\@undefined}
\makeatother

\begin{document}

\title{Electrical tuning of tin-vacancy centers in diamond}
\author{Shahriar Aghaeimeibodi}
\altaffiliation{These authors contributed equally to this work.}
\affiliation{E. L. Ginzton Laboratory, Stanford University, Stanford, CA 94305, USA}
% \thanks{These authors contributed equally to this work.}
\affiliation{shahriar@stanford.edu}

\author{Daniel Riedel}
\altaffiliation{These authors contributed equally to this work.}
\affiliation{E. L. Ginzton Laboratory, Stanford University, Stanford, CA 94305, USA}
\author{Alison E. Rugar}
\altaffiliation{These authors contributed equally to this work.}
\affiliation{E. L. Ginzton Laboratory, Stanford University, Stanford, CA 94305, USA}
\author{Constantin Dory}
\affiliation{E. L. Ginzton Laboratory, Stanford University, Stanford, CA 94305, USA}
\author{Jelena Vu\v{c}kovi\'c}
\affiliation{E. L. Ginzton Laboratory, Stanford University, Stanford, CA 94305, USA}

\begin{abstract}
% Group-IV color centers in diamond have attracted significant attention as a solid-state spin qubit due to their excellent optical and spin properties. In particular, large ground-state splitting allows the spin-state in tin-vacancy (SnV$^{\,\textrm{-}}$) center to maintain its coherence at temperatures above 1 K. The inversion-symmetry in the atomic structure of group-IV color centers significantly reduces the susceptibility of optical transitions to electric field fluctuations enabling their integration with nanostructures. However, this has not been directly probed (\co{needs rewording obviously}). In this letter, we demonstrate that the transition energy of SnV$^{\,\textrm{-}}$ center predominantly scales quadratically with an applied electric field. This observation directly confirms the robustness of SnV$^{\,\textrm{-}}$ emission to the electric field noise to the first order. We further show reversible tuning of the emission energy up to 800 MHz without a significantly broadening in the linewidth or decrease in the intensity.
% \\
% \begin{itemize}
%     \item[$\rightarrow$] Group IV good for quantum info .We need indistinguishable emitters
%     \item[$\rightarrow$] Tuning mechanism are either global or none efficient 
%     \item[$\rightarrow$]Stark effect has not been characterized 
%     \item[$\rightarrow$]We show 1.5 GHz (0.8 Ghz) shift 
%     \item[$\rightarrow$]We distinguish with heating
%     \item[$\rightarrow$]Strain in our sample
%     \item[$\rightarrow$]remaining linear shift
%     \item[$\rightarrow$]This could pave the way
% \end{itemize}

Group-IV color centers in diamond have attracted significant attention as solid-state spin qubits because of their excellent optical and spin properties. Among these color centers, the tin-vacancy (SnV$^{\,\textrm{-}}$) center is of particular interest because its large ground-state splitting enables long spin coherence times at temperatures above 1~K. However, color centers typically suffer from inhomogeneous broadening, which can be exacerbated by nanofabrication-induced strain, hindering the implementation of quantum nodes emitting indistinguishable photons. Although strain and Raman tuning have been investigated as promising techniques to overcome the spectral mismatch between distinct group-IV color centers, other approaches need to be explored to find methods that can offer more localized control without sacrificing emission intensity. Here, we study electrical tuning of SnV$^{\,\textrm{-}}$ centers in diamond via the direct-current Stark effect. We demonstrate a tuning range beyond 1.7~GHz. We observe both quadratic and linear dependence on the applied electric field. We also confirm that the tuning effect we observe is a result of the applied electric field and is distinct from thermal tuning due to Joule heating. Stark tuning is a promising avenue toward overcoming detunings between emitters and enabling the realization of multiple identical quantum nodes.
% Quantum networks based on optically interfaced quantum nodes require the homogeneous quantum emitters.
% While certain sample preparation and color center generation techniques may be employed to decrease inhomogeneous broadening of color centers down to $\sim 15$~GHz, tuning methods to overcome the remaining inhomogeneous broadening must be explored.
%Among these color centers, the tin-vacancy center is of particular interest because its large ground-state splitting enables long spin coherence times at temperatures above 1~K. 

\end{abstract}

\maketitle

\section{Introduction}
Group-IV color centers in diamond have emerged as promising candidates for the implementation of quantum networks\,\cite{Kimble2008,Wehner2018} because they retain their optical coherence when integrated with nanophotonic devices\,\cite{Sipahigil2016,Rugar2020b,Evans2016,Wan2020}. These color centers are inversion-symmetric and thus exhibit a vanishingly small permanent electric dipole moment, which mitigates the influence of electric field fluctuations and results in a high frequency stability of their optical transitions\,\cite{Thiering2018}. Most notably, the negatively charged silicon-vacancy (SiV$^{\,\textrm{-}}$) center in diamond has been used to implement several quantum information processing applications\,\cite{Evans2018, Nguyen2019,Bhaskar2020}. However, the relatively small ground-state splitting (GSS) of SiV$^{\,\textrm{-}}$ centers restricts the long spin coherence of this emitter to millikelvin temperatures. On the other hand, the negatively charged tin-vacancy (SnV$^{\,\textrm{-}}$) center comprises a heavier group-IV element in a split-vacancy configuration\,\cite{Wahl2020,Rugar2019, Rugar2020,Iwasaki2017,Gorlitz2020,Rugar2021}. The comparatively large GSS of SnV$^{\,\textrm{-}}$ centers facilitates long spin coherence times at liquid helium temperatures ($>$1~K) because of the reduced phonon-induced decoherence\,\cite{Trusheim2020}.

%The coupling of frequency-stable, negatively-charged silicon-vacancy (SiV$^{\,\textrm{-}}$) centers to diamond nanocavities enabled the engineering of photon-mediated interactions between two quantum emitters coupled to the same cavity\,\cite{Evans2018}. Furthermore, establishing an efficient nanophotonic interface to a long-lived nuclear spin memory\,\cite{Nguyen2019} memory-enhanced quantum communication was demonstrated\,\cite{Bhaskar2020}.
Quantum networks require multiple identical quantum nodes that emit indistinguishable photons. A significant challenge for solid-state quantum emitters is that each individual emitter experiences a slightly different strain environment, leading to an inhomogeneous distribution of their optical transition frequencies. Because of the aforementioned inversion symmetry, the spread of optical transition frequencies of group-IV color centers can be very narrow in a low-strain environment\,\cite{Rogers2014,Sipahigil2014}. Nevertheless, in diamond nanostructures, fabrication-induced strain produces a slightly larger inhomogeneous broadening (\textless~15~GHz)\,\cite{Evans2016}. Therefore, to achieve indistinguishable emission from distinct color centers, precise tuning of optical transition frequencies is essential. Raman\,\cite{Sipahigil2016,Sun2018} and magnetic field\,\cite{Evans2018} tuning have been employed to control the emitter frequency over ranges comparable to this spread. Those techniques, however, either strongly reduce the photon detection rate or cannot be applied locally on several emitters on the same chip to compensate their frequency detuning.
Electromechanical tuning offers a potential solution and has enabled the tuning of the transition frequency of waveguide-coupled SiV$^{\,\textrm{-}}$ centers by tens of GHz\,\cite{Machielse2019,Wan2020}. This technique, however, is limited to freestanding waveguide structures. Applying an electric field could potentially offer an alternative tuning mechanism through the Stark effect. However, because of the predicted first order insensitivity of transitions to electric fields, this approach has not been realized.

In this Letter, we investigate the electric field susceptibility of SnV$^{\,\textrm{-}}$ centers in diamond. 
We demonstrate reversible tuning of the transition wavelength by more than 1.7~GHz, which is $\sim$ 57 times the natural linewidth of $\sim$ 30~MHz\,\cite{Trusheim2020}. We measure linear and quadratic Stark effect coefficients for SnV$^{\,\textrm{-}}$ centers to be several orders of magnitude smaller than those of non-inversion-symmetric color centers. The remaining linear shift in some emitters may originate from the relatively high strain in our diamond sample, which we infer from the large distribution of transition frequencies ($\sim$ 270~GHz) of SnV$^{\,\textrm{-}}$ centers in this sample, as previously characterized in Ref.~\onlinecite{Rugar2019}. Furthermore, we perform careful control experiments to distinguish the tuning achieved through the Stark effect from that resulting from parasitic heating of the diamond sample. Our results, when combined with an initial pre-selection of emitters for nearby transition frequencies, could enable the realization of multi-emitter quantum information processing applications using diamond color centers.

\begin{figure*}[t]
\includegraphics[width=0.9\textwidth,]{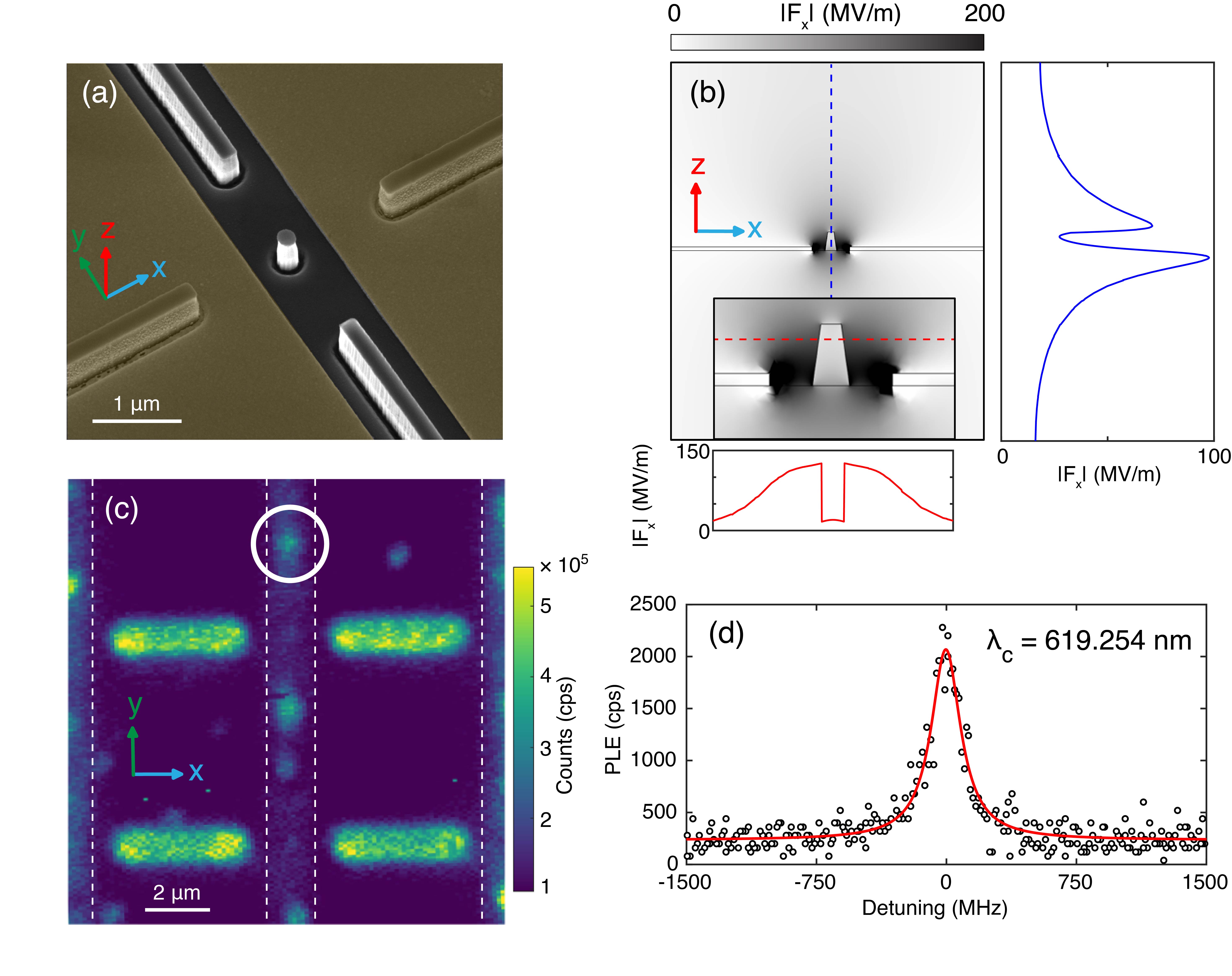}
\caption{Device fabrication and characterization. \textbf{(a)} Scanning electron microscope image of fabricated diamond structure and electrodes. Electrodes are highlighted by false-color gold. \textbf{(b)} Simulated electric field magnitude along the x-direction ($\left|F_x\right|$) for an applied voltage of 200 V across the electrodes. Inset shows close up of $\left|F_x\right|$ in mesa. Electric fields along the dashed blue and red lines are plotted in the panels to the right ($\left|F_x(z)\right|$) and below ($\left|F_x(x)\right|$), respectively. \textbf{(c)} Photoluminescence map of the device. One of the SnV$^{\,\textrm{-}}$ centers that we study in this work (E1) is circled in white. Edges of electrodes are demarcated by white dashed lines. \textbf{(d)} Photoluminescence excitation measurement performed on the C transition of the marked SnV$^{\,\textrm{-}}$ center in (c) at zero applied electric field. Lorentzian fit (red curve) to data (black circle) reveals a linewidth of ${194\pm12}$~MHz.}
\label{characterize}
\end{figure*}

\section{Fabrication and characterization}

We begin our fabrication process with an electronic-grade, single-crystal diamond plate from Element Six. The chip is cleaned in a boiling tri-acid (1:1:1 sulfuric/nitric/perchloric acids) solution. We then remove the top 300~nm of diamond with an oxygen (O$_2$) plasma etch. SnV$^{\,\textrm{-}}$ centers are generated via $^{120}$Sn$^+$ ion implantation (370~keV, $2\times10^{11}$~cm$^{-2}$) followed by vacuum annealing at 800$\degree\textrm{C}$ for 30~minutes and 1100$\degree\textrm{C}$ for 90~minutes. Based on Stopping Range of Ions in Matter (SRIM) simulations\,\cite{Ziegler2010}, we expect the depth of our SnV$^{\,\textrm{-}}$ centers to be $\sim90$~nm.

After generating SnV$^{\,\textrm{-}}$ centers in diamond, we fabricate nanopillars and mesas to easily identify single SnV$^{\,\textrm{-}}$ centers and to increase the photon extraction efficiency. We grow 200~nm of silicon nitride (Si$_x$N$_y$) via low-pressure chemical vapor deposition. We pattern square arrays of circles (pillars) separated by rectangles (mesas) in hydrogen silsesquioxane FOx-16 via electron-beam (e-beam) lithography. The circles range in diameter from 140~nm to 300~nm. The rectangles are $0.2~\mu\textrm{m}\times2.5~\mu\textrm{m}$. The pattern is transferred into the Si$_x$N$_y$ film by a SF$_6$, CH$_4$, and N$_2$ reactive ion etch. Using the Si$_x$N$_y$ layer as an etch mask, we perform an O$_2$ plasma etch to fabricate 500-nm tall diamond nanopillars and mesas. Finally, the etch mask is removed by soaking the sample in hydrofluoric acid.

We then fabricate parallel electrodes around a column in an array of nanopillars to produce an electric field along the $\left[100\right]$ direction. The 4-$\mu\textrm{m}$ wide electrodes are placed 1~$\mu\textrm{m}$ apart. The electrodes are created via metal liftoff. Poly(methyl methacrylate) (PMMA) is patterned via e-beam lithography. Then 5~nm of Ti followed by 30~nm of Au are deposited by e-beam evaporation. The remaining PMMA is lifted off in acetone. A scanning electron microscope image of the resulting diamond structures and metal electrodes is shown in Fig.~\ref{characterize}(a).

We use the finite element method (COMSOL) to simulate the electric field inside a pillar or a mesa. Fig.~\ref{characterize}(b) shows the simulated magnitude of the electric field along the x direction ($\left|F_x\right|$) when 200~V is applied across the electrodes. The inset is a close-up view of the field distribution around the mesa, indicating a lower magnitude in the mesa compared to the bulk region. The right panel in Fig.~\ref{characterize}(c) is a vertical line cut through the center of a diamond mesa, showing $\left|F_x\right|$ as a function of z position, while the bottom panel shows $\left|F_x\right|$ as a function of x at the estimated implantation depth. The electric field magnitude is a factor of three smaller at $\sim$ 90~nm below the top of the mesa, than at the surface of diamond for a bulk region. 

We perform the optical characterization of SnV$^{\,\textrm{-}}$ centers in this sample at $\sim5$~K in a cryostat (Montana Instruments Cryostation). Using a home-built scanning confocal microscope setup, we scan a 532-nm continuous-wave laser across our sample and collect the emission into the SnV$^{\,\textrm{-}}$ center zero-phonon line with a $(620\pm14)$-nm bandpass filter placed in front of a multi-mode fiber to acquire a photoluminescence (PL) map of the region (Fig.~\ref{characterize}(c)). 

%For the remainder of this paper, we will focus on the emitter circled in white in Fig.~\ref{characterize}(d).
At 5 K, the SnV$^{\,\textrm{-}}$ center has two dominant zero-phonon line transitions often referred to as the C and D transitions\,\cite{Iwasaki2017}. These transitions are centered around 620~nm, and in unstrained SnV$^{\,\textrm{-}}$ centers the C transition is typically $\sim$ 850 GHz higher in energy than the D transition. We perform a photoluminescence excitation (PLE) measurement on the C transition of the emitter circled in Fig.~\ref{characterize}(c), which we will refer to as E1, to characterize its resonant frequency and linewidth. We scan the wavelength of a tunable laser (MSquared SolsTiS and External Mixing Module) around the transition wavelength and collect the emission into the phonon sideband (PSB) with a 638-nm long-pass filter and a 700-nm short-pass filter placed before a multi-mode collection fiber. When the laser passes through resonance with the C transition, a peak in the PSB photon counts occurs. To this peak, shown in Fig.~\ref{characterize}(e), we fit a Lorentzian curve to find the center wavelength 619.254~nm and linewidth (194$\pm$12)~MHz.

\section{Voltage-Dependent Photoluminescence Excitation}
\begin{figure}[t]
\includegraphics[width=0.4\textwidth,]{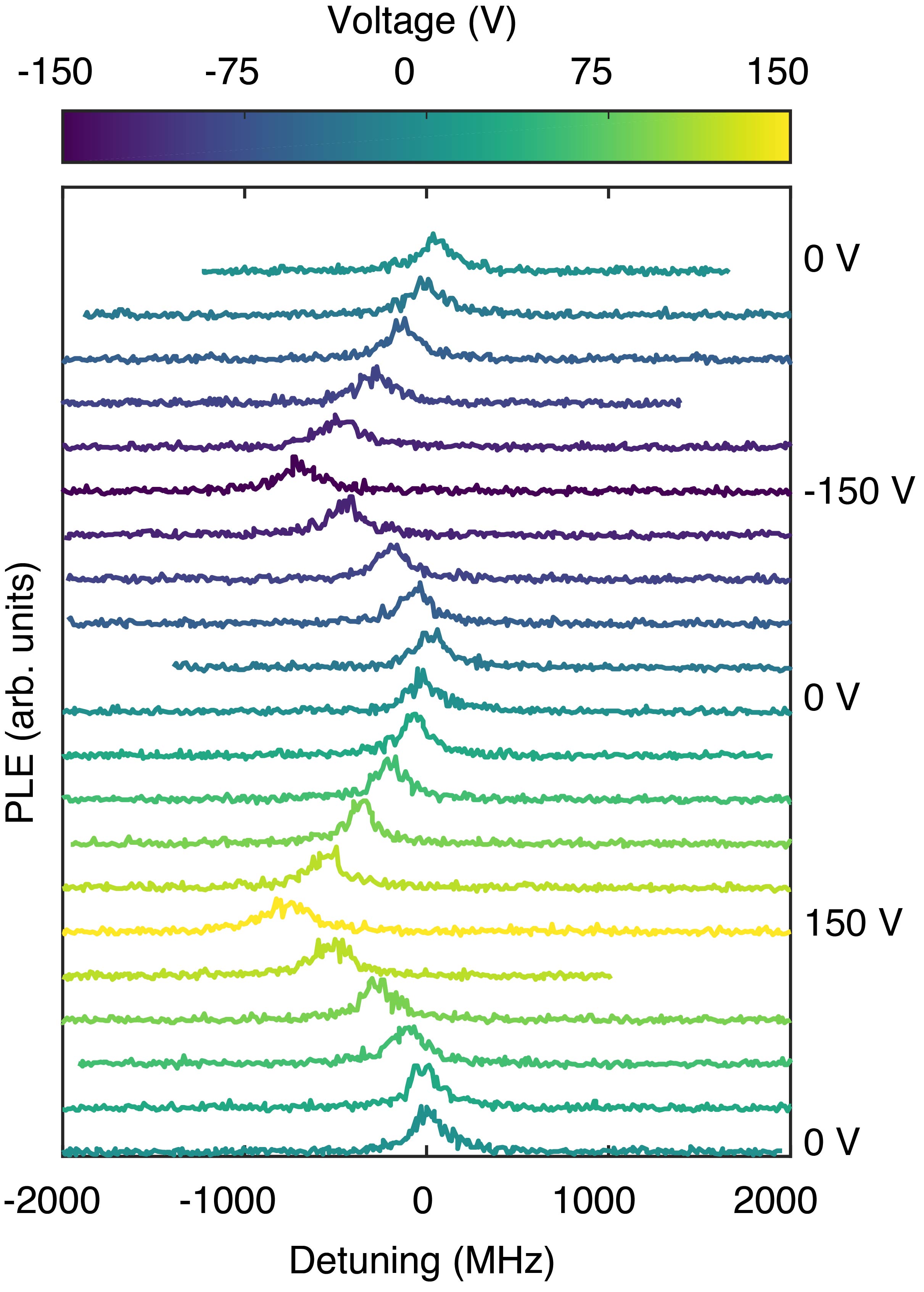}
\caption{PLE measurements acquired consecutively at different applied voltages. The PLE resonance reproducibly shifts as a function of applied voltage.}
\label{PLE}
\end{figure}

To investigate the behavior of the SnV$^{\,\textrm{-}}$ center in the presence of an electric field, we apply a direct-current (DC) voltage to the electrodes using a high-voltage power supply (Stanford Research Systems PS325). Fig.~\ref{PLE} shows consecutive PLE scans of the color center E1 as we vary the applied voltage. We observe a reversible redshift in the resonant frequency of the SnV$^{\,\textrm{-}}$ center for both polarities of the electric field. Repeatable tuning of the emission frequency confirms that there is no damage to the SnV$^{\,\textrm{-}}$ even at extremely high electric fields of $\sim$ 50 \textrm{MV/m}.

To gain a better understanding of the origin of the shift for single SnV$^{\,\textrm{-}}$ centers, we fit the PLE data in Fig.~\ref{PLE} to a Lorentzian function to extract the shift, linewidth, and the intensity of the signal as we vary the electric field. We approximate the electric field at the location of the emitter using the Lorentz approximation $F_\textrm{Local} = F_\textrm{ext}(\epsilon+2)/3$, where $F_\textrm{ext}$ is the field extracted from the COMSOL simulations and $\epsilon$ is the dielectric constant of diamond\,\cite{Tamarat2006}. %The lateral location of the emitter can be obtained using a Gaussian fit to the PL intensity map of the Fig.~\ref{characterize}(d).

\subsection{Quadratic shift}
\begin{figure}[htbp]
\includegraphics[width=0.4\textwidth,]{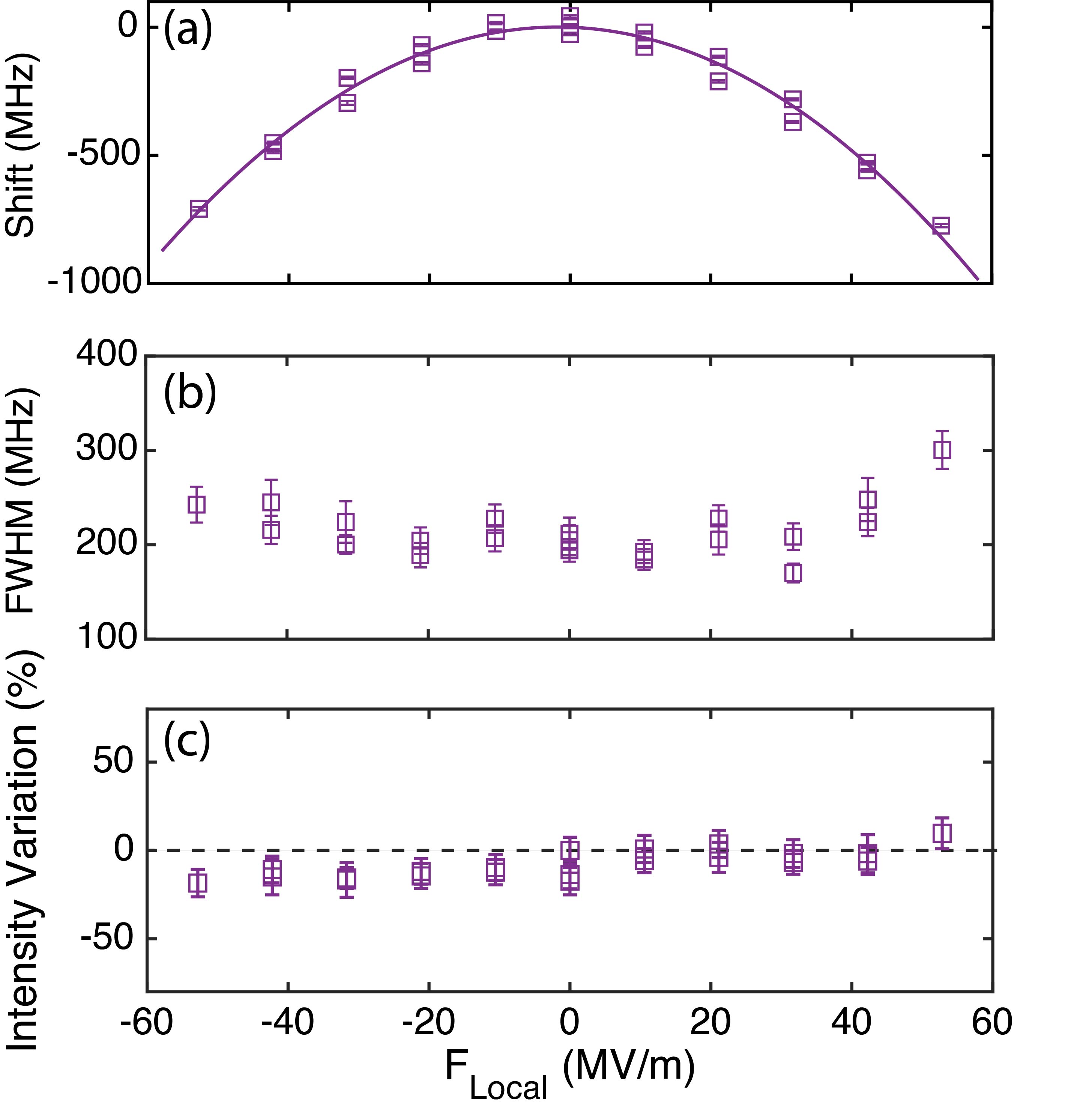}
\caption{Characterizing the Stark effect in a single SnV$^{\,\textrm{-}}$ center (E1) with predominantly quadratic shift. \textbf{(a)} Stark shift extracted from Lorentzian fits to the PLE data in Fig.~\ref{PLE} as a function of the applied electric field $F_\textrm{Local}$. The shifts show a mostly quadratic dependence on the applied field with $\Delta\mu = (1.9 \pm 1.1) \times 10^{-4}$ D and $\Delta\alpha /(4\pi\epsilon_0)$ = (3.28 $\pm$ 0.18) $\textrm{\AA}^3$ obtained from the fit. \textbf{(b)} Full width at half-maximum (FWHM) of the PLE signal of the same emitter as a function of $F_\textrm{Local}$. At high electric fields, the linewidth increases to $\sim$ 300~MHz from its original value $\sim$ 200~MHz value at zero field. \textbf{(c)} Integrated intensity of the PLE data as a function of $F_\textrm{Local}$. We observe $<\pm20\%$ variation in the integrated intensity of the PLE signal for the entire range of the electric field.}
\label{quadratic}
\end{figure}

Fig.~\ref{quadratic}(a) shows the extracted shift of the resonance frequency of the SnV$^{\,\textrm{-}}$ center E1 as a function of the applied electric field. We fit the extracted shift ($\Delta E$) in Fig.~\ref{quadratic}(a) to a quadratic function of the form $\Delta E = -\Delta \mu F_\textrm{Local} - 1/2 \Delta \alpha F_\textrm{Local}^2$ where $\Delta \mu$, and $\Delta \alpha$ are the change in dipole moment and polarizability between the excited and ground states. An inversion-symmetric defect such as the SnV$^{\,\textrm{-}}$ center is expected to have a vanishing $\Delta \mu$, making polarizability $\Delta \alpha$ the predominant coefficient. From the fit, we extract $\Delta \mu = (0.97 \pm 0.57)$~MHz/(MV/m) which corresponds to $(1.9 \pm 1.1) \times 10^{-4}$ D. This value is several orders of magnitude smaller than that of non-inversion-symmetric color centers such as nitrogen-vacancy (NV$^{\,\textrm{-}}$) centers in diamond\,\cite{Tamarat2006} and silicon vacancy (V$_\textrm{Si}$) centers in silicon carbide\,\cite{Lukin2020}. We also obtain $\Delta \alpha = (0.55 \pm 0.03)$~MHz/(MV/m)$^2$ which converts to a polarizability volume $\Delta\alpha /(4\pi\epsilon_0)$ = 3.28 $\pm$ 0.18 $\textrm{\AA}^3$. We note that this polarizability volume is four orders of magnitude smaller and has the opposite sign compared to that of NV$^{\,\textrm{-}}$ centers in diamond\,\cite{Tamarat2006}. Most emitters have a positive polarizibility because excited states tend to be more polarizable than ground states\,\cite{Tamarat2006}.
% , since excited states are generally more polarizable than ground states\,\cite{Tamarat2006}.

. %We repeat this measurement for multiple SnV$^{\,\textrm{-}}$ centers between the electrodes and report the extracted parameters in Table.~\ref{table1}. 

%%% Fig3, details of the field dependent shift, origin of shift

%%% Fig4, details of the field dependent shift, origin of shift for the linear emitter
\begin{figure}[htbp]
\includegraphics[width=0.4\textwidth,]{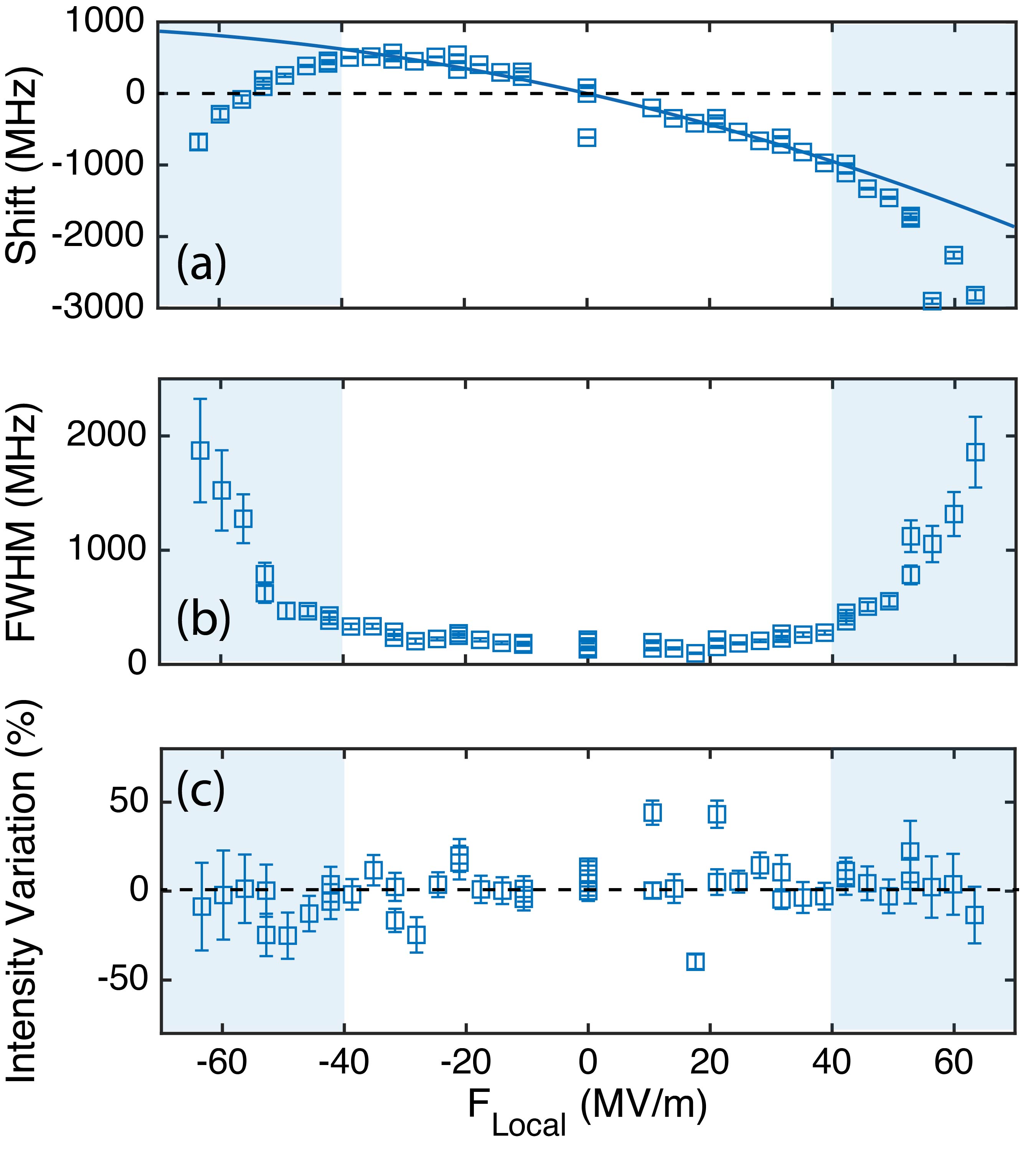}
\caption{Characterizing the Stark effect in a single SnV$^{\,\textrm{-}}$ center (E2) with predominantly linear shift. \textbf{(a)} Stark shift extracted from Lorentzian fits to the voltage-dependent PLE spectra of E2. The shifts show a mostly linear dependence on the applied field with $\Delta\mu = (3.9\pm0.4)\times10^{-3}$~D and $\Delta\alpha/(4\pi\epsilon_0) = (1.19\pm0.89)~\textrm{\AA}^3$ obtained from the fit. \textbf{(b)} FWHM of the PLE signal of the same emitter as a function of electric field $F_\textrm{Local}$. At high electric fields (shaded area), the linewidth rapidly increases. We attribute this significant line broadening to Joule heating through leakage current, as we will discuss later in the text. \textbf{(c)} Integrated intensity of the PLE data as a function of $F_\textrm{Local}$. We observe $<\pm45\%$ variation in the integrated intensity of the PLE signal for the entire range of the electric field.} %(d) FWHM as a function of the extracted shift in the PLE signal for three different tuning methods. Purple (Blue) squares correspond to the Stark tuning of the emitter E1 (E2). Yellow squares show the data for an emitter E$_\textrm{out}$ located outside the electrodes. Black circles represent the tuning of emitter E1 obtained by heating the sample stage. Similar slope of the black circles and yellow squares confirm that they both originate from heating and are fundamentally different than the shift obtained by the Stark effect. The yellow solid line is a linear fit indicating that in the heating regime, linewidth scales linearly with the shift.}
\label{linear}
\end{figure}

Fig.~\ref{quadratic}(b) shows the linewidth of the SnV$^{\,\textrm{-}}$ center obtained from the full width at half-maximum (FWHM) of the Lorentzian fits to the PLE data as a function of applied electric field. We observe an increase in the linewidth from the zero-field value of $\sim$ 200~MHz to 300~MHz for the highest electric field. We attribute this increase in linewidth to Joule heating because of the leakage current, as we discuss later. Fig.~\ref{quadratic}(c) displays the variation in integrated intensity of the PLE signal as a function of the applied electric field. We observe less than $\pm$20\% variation in the integrated intensity of the signal for a large range of applied electric fields, indicating the potential of this technique as a means to tune SnV$^{\,\textrm{-}}$ centers into resonance with each other.
\begin{figure*}[htbp]
\includegraphics[width=0.75\textwidth,]{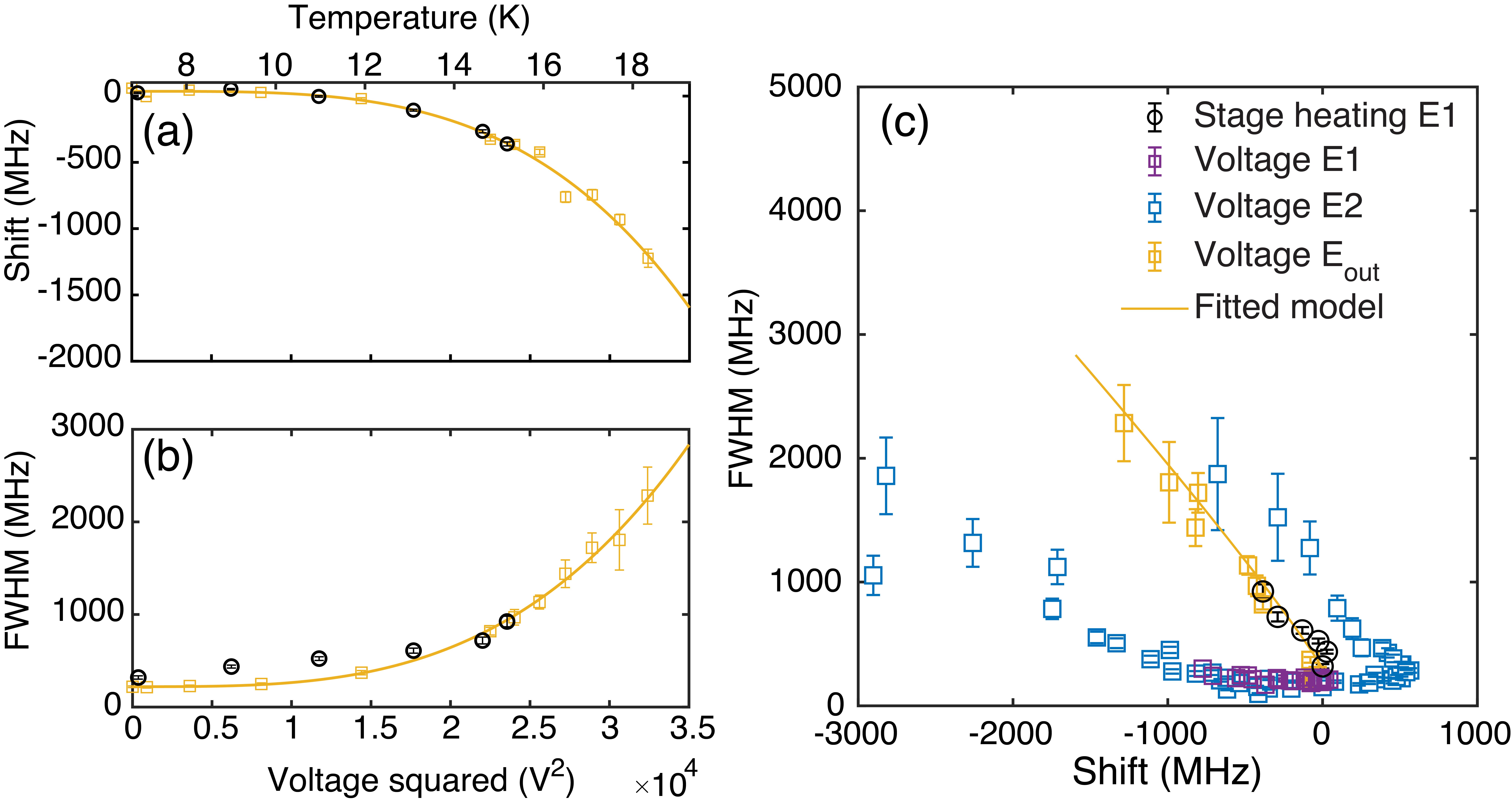}
\caption{A comparison between Stark and temperature tuning. \textbf{(a)} Effect of Joule heating on resonant frequency shift of an emitter E$_\textrm{out}$ located outside the electrodes. A power law model (yellow curve) is fit to the data (yellow squares), 
indicating a dependence on temperature ($T\propto V^2$), $T^{3.6\pm0.7}$. Black circles correspond to the temperature-dependent shift for E1 obtained from heating the sample. \textbf{(b)} Effect of Joule heating on the FWHM of E$_\textrm{out}$. A power law model 
(curve) is fit to the data (squares) with a $T^{3.3\pm0.4}$ dependence on temperature. Black circles correspond to the temperature-dependent FWHM for E1 obtained from heating the sample. \textbf{(c)} FWHM as a function of the 
extracted shift in the PLE signal for three different tuning methods. Purple (Blue) squares correspond to the Stark tuning of the emitter E1 (E2). Yellow squares show the data for 
E$_\textrm{out}$. Black circles represent the tuning of 
emitter E1 obtained by heating the sample stage. The similar behaviors of the black circles and yellow squares confirm that they both originate from heating and are fundamentally different from the shift obtained by the Stark effect. The yellow solid line is the curve resulting from the fitted models from panels (a) and (b) plotted against each other.}
\label{joule_heating}
\end{figure*}
%% Table for other emitters
 \begin{table*}[htbp]
 \centering
 \begin{tabular}{ |c |c |c| c| c| c|c| c|}
 \hline
 	 No. & $\lambda_\textrm{C}$ (nm) & Linewidth (MHz) & $\Delta\mu$ (10$^{-3}$D) & $\Delta\alpha/(4\pi\epsilon_0) (\textrm{\AA}^3)$ & GSS (GHz) & Shift (GHz) & Type \\
 	\hline
 	E1 & 619.254 & 194 $\pm$ 12 & 0.19 $\pm$ 0.11 & 3.28 $\pm$ 0.18 & 819.6& 0.82 & Quadratic \\
 	\hline
 	E2 & 619.236 & 132 $\pm$ 6 & 3.9 $\pm$ 0.4 & 1.19 $\pm$ 0.89 & 982.9&1.68 & Linear\\ 
 	\hline
 	E3 & 619.255 & 126 $\pm$ 6 & -0.5 $\pm$ 0.5 & 3.8 $\pm$ 1.1 & 947.7& 0.59 & Quadratic\\
 	\hline
 	E4 & 619.500& 160 $\pm$ 9 & -8.2 $\pm$ 1.8 & 5.36 $\pm$ 4.05 & 989.4& 1.73 & Linear\\
 	\hline
 \end{tabular}
 \caption{Linear and quadratic Stark effect coefficients of single SnV$^{\,\textrm{-}}$ centers. $\lambda_\textrm{C}$ (wavelength of C transition) and linewidth are obtained from Lorentzian fits to the PLE spectra at zero electric field. $\Delta\mu$ and $\Delta\alpha$ are Stark coefficients resulting from the same quadratic fit as in Fig.~\ref{quadratic}(a). The energy difference between the C and D transitions measured through PL spectroscopy gives the GSS. We define the useful shifting range as the largest tuning range achievable without broadening the linewidth by more than a factor of 2. Transition frequencies of E1 and E3 are $<1$ GHz apart.}
 \label{table1}
 \end{table*}
\subsection{Linear shift}
We study the electric-field dependence of transition frequency for several other SnV$^{\,\textrm{-}}$ centers. Although a quadratic shift is expected because of the inversion-symmetric structure of this color center, some emitters exhibit a mostly linear shift in transition frequency as a function of applied electric field.
Figs.~\ref{linear}(a)-(c) respectively display the extracted shift, linewidth, and intensity variation of PLE spectra for an emitter with mostly linear shift (E2). At local electric field magnitudes greater than 40~MV/m (shaded blue regions), we observe a redshift in the transition frequency and a sharp increase in the SnV$^{\,\textrm{-}}$ linewidth regardless of the field polarity. We attribute this linewidth broadening to Joule heating through leakage current. We observed that the onset voltage for the Joule heating decreased over time, indicating a possible damaged or burned area on the electrodes. In order to extract the Stark parameters without distortions caused by heating, we excluded the data in the shaded region for the fit. The solid blue line in Fig.~\ref{linear}(a) is a fit to the same model as in Fig.~\ref{quadratic}(a) with extracted parameters $\Delta\mu = (3.9\pm0.4)\times10^{-3}$~D and $\Delta\alpha/(4\pi\epsilon_0) = (1.19\pm0.89)~\textrm{\AA}^3$. This $\Delta\mu$ is more than an order of magnitude larger than that of E1 studied in Fig.~\ref{quadratic} and could be due to broken inversion symmetry in a strained SnV$^{\,\textrm{-}}$ center. The larger GSS in E2 ($\sim$ 983~GHz) compared to E1 ($\sim$ 820~GHz) is another indication of a higher strain environment for E2\,\cite{Meesala2018}. We repeat this measurement for multiple SnV$^{\,\textrm{-}}$ centers between the electrodes and report the extracted parameters in Table~\ref{table1}.

\subsection{Joule heating}
To confirm that the observed Stark shift is not related to Joule heating of the color center because of the leakage current, we perform two additional control experiments: we tune a SnV$^{\,\textrm{-}}$ center either (1) by applying a voltage across the electrodes while measuring an emitter outside of the electrode region (E$_\textrm{out}$) or (2) by heating the sample and studying E1. We extract the linewidth and resonance frequency of the SnV$^{\,\textrm{-}}$ centers through PLE measurements. 
In our first control experiment, we study another emitter E$_\textrm{out}$ that is located outside of the electrode region. Because E$_\textrm{out}$ is sufficiently far from the electrodes, the DC electric field vanishes and any shift in frequency for this emitter should be due to leakage current, Joule heating, and the high thermal conductivity of diamond. In Figs.~\ref{joule_heating}(a)~and~(b), we plot the shift and FWHM as functions of the square of the voltage applied across the electrodes ($V^2$). With Joule heating, the induced temperature change is proportional to the applied voltage squared ($T\propto V^2$), so we use these data to characterize the temperature-dependent behavior of E$_\textrm{out}$. We fit a power law model to the $E_\textrm{out}$ data of Figs.~\ref{joule_heating}(a)~and~(b). From these fits, we find that the thermally induced shift is proportional to $T^{3.6\pm0.7}$ and FWHM $\sim T^{3.3\pm0.4}$. This temperature dependence of frequency and FWHM is consistent with previous studies of SiV$^{\,\textrm{-}}$ centers\,\cite{Jahnke2015}.
For the second control experiment, we tune the energy of the SnV$^{\,\textrm{-}}$ center studied in Fig.~\ref{quadratic} (E1) by heating the cryostat up to $\sim$ 20~K. We plot the detuning and FWHM data of E1 as a function of temperature in Figs.~\ref{joule_heating}(a)~and~(b). We use the extracted shift of the heated E1 and E$_\textrm{out}$ to calibrate the proportionality constant between $V^2$ and $T$.

In Fig.~\ref{joule_heating}(c), we plot FWHM as a function of frequency shift for our two control experiments as well as E1 (Fig.~\ref{quadratic}) and E2 (Fig.~\ref{linear}). We also plot the modeled FWHM against the modeled shift from Figs.~\ref{joule_heating}(a) and (b) as a solid yellow line. The data from the two temperature-tuning control experiments are consistent with each other and exhibit a vastly different behavior from the electrically tuned E1. Furthermore, we can see that at large detunings the slope of the E2 data increases dramatically, approaching the behavior of the thermally tuned emitters. 
These comparisons prove that while leakage current through electrodes can, in principle, cause Joule heating and broadening of the PLE data, the Stark shifts presented in Figs.~\ref{quadratic}(a) and \ref{linear}(a) are not influenced by this effect until very high local fields exceeding 40 MV/m.

% Yellow squares in Fig.~\ref{joule_heating}(c) show the dependence of linewidth on the shift for E$_\textrm{out}$. The trend is comparable to thermal tuning data and vastly contrasting from the Stark tuning of E1.

% In Fig.~\ref{linear}(d), we plot the linear fit to the FWHM versus shift data of E$_\textrm{out}$. We include lines parallel to the fit line to show that the behavior of E2 becomes dominated by thermal effects at beyond 40 MV/m.

% Black circles in Fig.~\ref{linear}(d) show the FWHM as a function of the extracted shift. For comparison, purple (blue) squares correspond to the voltage tuning of the emitter E1 (E2). We observe that the broadening of the linewidth for a similar shift in resonance frequency is significantly higher for temperature tuning confirming minimal Joule heating in our Stark tuning experiment. 

\section{Conclusions}

We have examined the response of the frequency of the C transition of SnV$^{\,\textrm{-}}$ centers in diamond to externally applied electric fields. We can shift the transition frequency by more than 1.7~GHz without introducing a significant broadening of the transition linewidth. We investigate several SnV$^{\,\textrm{-}}$ centers and observe both linear and quadratic dependencies of the shift on the applied electric field. While a quadratic dependence is expected based on the defect’s inversion symmetry, we attribute the linear shift to a small, strain-induced dipole moment. We find changes in dipole moment and polarizability volume between the excited and ground states of up to $\Delta\mu=-8.2\times10^{-3}$~D and $\Delta\alpha/(4\pi\epsilon_0) = 5.4~\textrm{\AA}^3$, respectively. Furthermore, we confirm that the observed shift is due to the Stark effect and is distinct from any heating effects by comparing the linewidth versus frequency shift of the emitter when a voltage is applied to that when the stage is heated.
%\co{we need to include linear shift here, what about the emitter with \textgreater GHz linear shift?}

Stark tuning can be used to tune the optical frequency of remote emitters and to control the degree of indistinguishability in two-photon interference experiments of pre-selected emitters\,\cite{Bernien2012,Sipahigil2012, Lettow2010,Patel2010,Flagg2010}. 
 This technique, which recently enabled the first demonstration of linking three remote NV$^{\,\textrm{-}}$ centers in a quantum network\,\cite{Pompili2021}, may also be applied to group-IV color centers.
Group-IV color centers have have very narrow inhomogeneous distribution in bulk\,\cite{Rogers2014,Sipahigil2014} and, because of their inversion symmetry, can also have relatively narrow inhomogeneous distributions in nanostructures (\textless~15~GHz)\,\cite{Evans2016}. Nanophotonic devices hosting two SiV$^{\,\textrm{-}}$ centers with a sub-GHz difference between their transition frequencies have been reported\,\cite{Evans2018}, which is within our Stark tuning range. On this chip, we have also observed sub-GHz frequency detuning between two SnV$^{\,\textrm{-}}$ centers (see Table~\ref{table1}). The 1.7~GHz tuning range that we demonstrate in this Letter is large enough to overcome modest detunings between color centers and enable multi-emitter experiments.
 
 The electric-field dependence of group-IV color centers can be harnessed in other schemes. Modulated electric fields have been proposed as an alternative way to overcome the inhomogeneous broadening of emitters\,\cite{Lukin2020,Trivedi2020}. This approach would be particularly important for overcoming detunings between closely spaced emitters, for example if the emitters are in the same nanophotonic cavity\,\cite{Lukin2020,Trivedi2020}. Spectral diffusion presents a challenge when working with group-IV color centers, which require resonant drive for optical spin initialization and control schemes. Feedback-based electric field tuning constitutes a tool for the dynamic stabilization of optical transition frequency of emitters\,\cite{Acosta2012}.

%By applying multiaxis electric fields using a four-port electrode configuration the transverse components of intrinsic strain could be compensated to restore the symmetry of the system\,\cite{Bassett2011}. (\co{maybe we should not talk about this to give away the idea})

%\co{With preselection - We already have very close emitters, XXX detuning}
All of these applications of Stark effect-based tuning can be made into more practical options by increasing the achievable Stark tuning range. The Stark tuning range can be further expanded by changing the electrode configuration to increase the field experienced by the emitter. Fabricating electrodes with a narrower gap between them would increase the applied field. Embedding the SnV$^{\,\textrm{-}}$ center below the plane of the electrodes rather than in a nanopillar would increase the field experienced by the emitter a factor of 3.
Furthermore, the tuning range can be increased by reducing the voltage-induced heating which allows the application of larger voltages.
%The main source of heating in our device is the leakage current between the electrodes. This heating could potentially be mitigated by reducing the surface conductivity of the diamond via chemical treatment and by adding an oxide layer to the electrodes\co{(refs)}.
% Due to the quadratic dependence of the Stark shift these improvements might enable tuning ranges comparable to the inhomogeneous distribution of transition frequencies in nanostructures while introducing only modest spectral diffusion.

% \co{Spectral diffusion presents a challenge when working with group-IV color centers, which require resonant drive for optical spin initialization and control schemes. Feedback-based electric field tuning constitutes a tool for the dynamic stabilization of optical transition frequency of emitters\,\cite{Acosta2012}. 
% % Modest spectral diffusion in nanostructures can also be overcome by Purcell-broadening the optical linewidth via coupling to nanophotonic resonators\,\cite{Sipahigil2016,Bhaskar2020, Rugar2021}.
% }

% \co{ Stark tuning has been used to tune the optical frequency of remote emitters and allowed controlling the degree of indistinguishability in two-photon interference experiments of pre-selected emitters\,\cite{Bernien2012,Sipahigil2012, Lettow2010,Patel2010,Flagg2010}. 
%  This technique recently enabled the first demonstration of linking three remote NV centers into a quantum network\,\cite{Pompili2021}.}
 
In addition to contributing to a deeper understanding of the basic properties of SnV$^{\,\textrm{-}}$ centers, our results pave the way for multi-emitter experiments based on group-IV color centers harnessing Stark shift tuning.

During the preparation of this manuscript, we became aware of a similar, very recent work\,\cite{DeSantis2021}.

\begin{acknowledgments}
We are grateful to Daniil Lukin for his experimental assistance to this work. This work is financially supported by Army Research Office (ARO) (award no. W911NF-13-1-0309); National Science Foundation (NSF) RAISE TAQS (award no. 1838976); Air Force Office of Scientific Research (AFOSR) DURIP (award no. FA9550-16-1-0223). S.A. acknowledges support from Bloch postdoctoral fellowship in quantum science and engineering from Stanford Q-FARM. D.R. acknowledges support from the Swiss National Science Foundation (Project P400P2\_194424). A.E.R. acknowledges support from the National Defense Science and Engineering Graduate (NDSEG) Fellowship Program, sponsored by the Air Force Research Laboratory (AFRL), the Office of Naval Research (ONR) and the Army Research Office (ARO). C.D. acknowledges support from the Andreas Bechtolsheim Stanford Graduate Fellowship (SGF) and the Microsoft Research PhD Fellowship. Part of this work was performed at the Stanford Nanofabrication Facility (SNF) and the Stanford Nano Shared Facilities (SNSF), supported by the National Science Foundation under award ECCS-2026822.\end{acknowledgments}

%\bibliography{SnVStark}
%\bibliography{bibliography}

\end{document}